\documentclass[prd,nofootinbib,showpacs,preprintnumbers,amsmath,amssymb]{revtex4}

\usepackage{graphicx}
\usepackage{dcolumn}
\usepackage{bm}

\def\bear{\begin{eqnarray}}
\def\ear{\end{eqnarray}}

\begin{document}

\begin{flushright}
YITP-13-52
\end{flushright}

\title{New Geometric Transition as Origin of Particle Production in Time-Dependent Backgrounds}

\author{Sang Pyo Kim}\email{sangkim@kunsan.ac.kr}
\affiliation{Department of Physics, Kunsan National University, Kunsan 573-701, Korea\footnote{Permanent address}}
\affiliation{Yukawa Institute for Theoretical Physics, Kyoto University, Kyoto 606-8502, Japan}

\begin{abstract}
By extending the quantum evolution of a scalar field in time-dependent backgrounds to the complex-time plane and transporting
the in-vacuum along a closed path, we argue that the geometric transition from the simple pole at infinity determines
the multi-pair production depending on the winding number. We apply the geometric transition to Schwinger mechanism
in the time-dependent vector potential for a constant electric field and to Gibbons-Hawking particle production
in the planar coordinates of a de Sitter space.
\end{abstract}

\date{\today}

\maketitle

The interaction of quantum field with a background gauge field or spacetime can produce particles as a nonperturbative quantum effect.
Schwinger mechanism is pair production by a constant electric field, which provides enough energy to separate charged pairs from the Dirac sea \cite{Schwinger}. Hawking radiation is the emission of particles from vacuum fluctuations,
which are separated by the horizon of a black hole \cite{Hawking}. Recently Hawking radiation has been interpreted as
quantum tunneling of virtual pairs near the horizon of the black hole \cite{Parikh-Wilzcek}. Also it has been known for long that an expanding spacetime produces particles \cite{Parker} and de Sitter (dS) radiation has a thermal distribution \cite{Gibbons-Hawking}.

In quantum field theory the out-vacuum of a quantum field may differ from the in-vacuum through the interaction with a background field
and  may be expressed as multi-particle states of the in-vacuum via the Bogoliubov transformation \cite{DeWitt75}.
In the in-out formalism the scattering matrix between the in-vacuum and the out-vacuum determines the probability
for the in-vacuum to remain in the out-vacuum, which in turn is given by the pair-production rate for bosons or fermions \cite{DeWitt03}.
The exact vacuum persistence and pair production requires the solution of the quantum field in the background field. With limited knowledge of exact solutions,
approximation scheme proves a practical approach or provides an intuitive understanding for pair production. Various approximation schemes have been proposed for Hawking radiation
\cite{VAD}.

Each Fourier mode of a scalar field in the time-dependent vector potential for a constant electric field and in the planar coordinates of a dS space
describes the scattering problem over a potential barrier.
In the phase-integral method Schwinger pair-production rate is determined by the action integral among quasi-classical turning points in the complex plane of time or space \cite{Kim-Page07,Dumlu-Dunne}. In particular, the action integral has been proposed as the contour integral in the complex plane for Schwinger mechanism \cite{Kim-Page07} and for Hawking radiation \cite{Kim08}. Furthermore, the Stokes lines and anti-Stokes lines for more than one pair of quasi-classical turning points distinguish boson and fermion pair production \cite{Dumlu-Dunne} and the dimensionality of particle production in dS spaces \cite{Kim10}. The complex analysis has been used in connection with particle production \cite{Vakkuri-Kraus,Srinivasan-Padmanabhan}, the instanton action \cite{Kim-Page02,Kim-Page06}, and
the worldline instanton \cite{Dunne-Schubert,DWGS}.

In this paper we propose the geometric transition of the Hamiltonian in the complex-time plane as a new interpretation of particle production in time-dependent backgrounds. It has been observed that a time-dependent Hamiltonian can have the geometric transition in the complex-time plane \cite{Hwang-Pechukas,JKP,Kim12}.
In the functional Schr\"{o}dinger picture each Fourier mode of quantum field in a time-dependent background has the Hamiltonian with time-dependent frequency and/or mass. We argue that the evolution of the in-vacuum along a complex closed path of non-zero winding number leads to the geometric transition from the simple pole at infinity and results in particle production, in strong contrast with the trivial real-time evolution without level-crossings. The evolution of the in-vacuum along a path in the complex-time plane is reminiscent of the closed-time path integral in the in-in formalism \cite{DeWitt03}.

A complex scalar field with mass $m$ and charge $q$ in a constant electric field in the (d+1)-dimensional Minkowski spacetime has the Fourier-decomposed, time-dependent Hamiltonian [in units of $c = \hbar =1$]
\begin{eqnarray}
H (t) = \int d^d {\bf k} \Bigl[\frac{1}{2} \pi_{\bf k}^{\dagger} \pi_{\bf k} + \frac{1}{2} \omega_{\bf k}^2 \phi^{\dagger}_{\bf k} \phi_{\bf k}  \Bigr], \label{qed ham}
\end{eqnarray}
where
\begin{eqnarray}
\omega_{\bf k}^2 (t) = m^2 + {\bf k}_{\perp}^2 + (k_{\parallel} + qEt)^2. \label{qed freq}
\end{eqnarray}
Here ${\bf k}_{\perp}$ and $k_{\parallel}$ are the transverse and longitudinal momenta and the vector potential is $A_{\parallel} (t) = - Et$, which provides
a time-dependent background. The complex scalar field is equivalent to two real scalar fields: one for particle and the other for antiparticle. In the planar coordinates of the (d+1)-dimensional dS space
\begin{eqnarray}
ds^2 = -dt^2 + e^{2H_{HC} t} d{\bf x}^2_d,
\end{eqnarray}
where $H_{HC}$ is the Hubble constant, a massive real scalar field has the time-dependent Hamiltonian
\begin{eqnarray}
H (t) = \int d^d {\bf k} \Bigl[ \frac{1}{2 M(t)} \pi_{\bf k}^2 + \frac{M(t)}{2} \omega_{\bf k}^2 (t) \phi_{\bf k}^2 \Bigr], \label{ds ham}
\end{eqnarray}
where
\begin{eqnarray}
M(t) = e^{dH_{HC}t}, \quad \omega_{\bf k}^2 (t) = m^2 + \frac{{\bf k}^2}{e^{2H_{HC} t}}. \label{ds freq}
\end{eqnarray}
In the functional Schr\"{o}dinger picture, the quantum state obeys the time-dependent Schr\"{o}dinger equation
\begin{eqnarray}
i \frac{\partial}{\partial t} \vert \Psi (t) \rangle = \hat{H} (t) \vert \Psi (t) \rangle, \quad \vert \Psi (t) \rangle = \prod_{\bf k} \vert \Psi_{\bf k} (t) \rangle.
\end{eqnarray}

For the purpose of this paper, it is sufficient to consider the time-dependent oscillator defined along the real-time axis as
\begin{eqnarray}
H (t) = \frac{1}{2M(t)} p^2 + \frac{M(t)}{2} \omega^2 (t) q^2, \quad (\omega (t) > 0).
\end{eqnarray}
In the real-time evolution, the annihilation and creation operators
\begin{eqnarray}
\hat{a} (t) = \sqrt{\frac{M(t) \omega (t)}{2}} \hat{q} + \frac{i}{\sqrt{2M(t) \omega(t)}} \hat{p}, \quad
\hat{a}^{\dagger} (t) = \sqrt{\frac{M(t) \omega (t)}{2}} \hat{q} - \frac{i}{\sqrt{2M(t) \omega(t)}} \hat{p}
\end{eqnarray}
diagonalize the Hamiltonian as
\begin{eqnarray}
\hat{H} (t) = \omega (t) \Bigl(\hat{a}^{\dagger} (t) \hat{a} (t) + \frac{1}{2} \Bigr). \label{num ham}
\end{eqnarray}
Thus, an initial state
\begin{eqnarray}
\vert \Psi(t) \rangle = \hat{U} (t, t_0)  \vert \Psi(t_0) \rangle,
\end{eqnarray}
evolves by the evolution operator, which can be given by the time-ordered integral or product integral \cite{Dollard-Friedman}
\begin{eqnarray}
\hat{U} (t, t_0) = {\rm T} \exp \bigl[- i \int_{t_0}^{t} \hat{H} (t') dt' \bigr] = \prod_{t_0}^{t} \exp \bigl[- i
\hat{H} (t') dt' \bigr].
\end{eqnarray}
In terms of the number states (\ref{num ham}), the evolution operator can be further written as \cite{Kim12}
\begin{eqnarray}
\hat{U} (t, t_0) = \Phi^T (t) {\rm T} \exp \bigl[- i \int_{t_0}^{t} \bigl(
H_D (t') - A^T (t') \bigr) dt' \bigr] \Phi^*(t_0),
\end{eqnarray}
where $H_{D} (t)$  and  $\Phi(t)$ denote the diagonal matrix and the column vector, respectively,
\begin{eqnarray}
H_{D} (t) = \omega (t) \begin{pmatrix}
\frac{1}{2} & & & &   \\
& & \ddots & &\\
& & & n+ \frac{1}{2} &\\
& & & & \ddots
 \end{pmatrix}, \quad \Phi (t) = \begin{pmatrix}
 \vert 0, t \rangle   \\
\vdots \\
 \vert n, t \rangle \\
\vdots
 \end{pmatrix}, \label{h mat}
\end{eqnarray}
and $A(t)$ is the induced vector potential
\begin{eqnarray}
A (t) = i \Phi^* (t) \frac{\partial \Phi^T (t)}{\partial t}  = i \frac{\dot{\omega} (t)}{4 \omega (t)} \bigl(\sqrt{n(n-1)}\delta_{m n-2} - \sqrt{(n+1)(n+2)} \delta_{m n+2} \bigr).
\end{eqnarray}
Here and hereafter overdots denote derivatives with respect to the real or complex time.
Hence $\hat{U} (t_0, t_0)$ from $t_0$ to any future time $t$ and back to $t_0$ along the real-time axis becomes unity since $H_D (t)$ and $A(t)$ do not have any singularity due to $\omega (t) > 0$, which is the case of charged scalars in a time-dependent vector potential or real scalars in a dS space. Note that the path from $t_0$ to $t_0$ along the real-time axis is a loop of zero-winding number, which will be denoted as $C^{(0)} (t_0)$ with the base point $t_0$. In other words, $\hat{U} (C^{(0)}(t_0)) = I$ in the complex-time plane and the scattering amplitude between the in-vacuum and the transported in-vacuum is unity along the real-time axis
\begin{eqnarray}
\langle 0,  C^{(0)}(t_0) \vert 0, t_0 \rangle = 1. \label{real-time scat}
\end{eqnarray}
The in-in formalism thus becomes trivial as long as the real time is concerned.

However, a time-dependent Hamiltonian, provided that it has a level-crossing in the complex-time plane, leads to the geometric transition amplitude, which is responsible for an exponential decay of the initial state and the transition to other states \cite{Hwang-Pechukas,JKP}. In a similar manner, let us extend the Hamiltonian $H(t)$  to the complex-time plane and assume that $H(z)$ is analytic and the orthonormality $\langle m, z \vert n, z \rangle = \delta_{mn}$ holds. In a properly chosen Riemann sheet in the complex-time plane, the frequencies (\ref{qed freq}) and (\ref{ds freq}) have two branch points of the form
\begin{eqnarray}
\omega (z) = f(z) \sqrt{(z- z_0) (z- z_0^*)}, \label{com freq}
\end{eqnarray}
where $f(z)$ is an analytic function. Though the complex frequency (\ref{com freq}) has level-crossings in the whole complex plane, we cut two branch lines from $z_0$ and $z_0^*$ as shown in Fig. 1, which make $\omega(z)$ analytic in the proper Riemann sheet.
Defining $\tilde{H} (z) := H_{D} (z) - A^T (z)$, we notice that the matrix-valued $\tilde{H} (z)$ does not commute with $\tilde{H} (z')$ in general
for $z \neq z'$ unless $\omega^2 (z') \dot{\omega}(z) = \omega^2 (z) \dot{\omega}(z')$.
Then, without level-crossings, all the time-ordered integrals of $\hat{H}(z)$ in the complex-time plane along closed paths of the same winding number and with the same base point $t_0$ on the real-time axis are equal to each other \cite{Dollard-Friedman}
\begin{eqnarray}
{\rm T}  \exp \bigl[-i \oint_{C_{I} (t_0)} \tilde{H} (z) dz \bigr] = {\rm T}  \exp \bigl[-i \oint_{C_{II} (t_0)} \tilde{H} (z) dz \bigr].
\end{eqnarray}
Hence the time integral in the complex-time plane is independent of paths and depends only on the homotopy class
of winding numbers. In fact, the path in the left panel of Fig. 1 has the winding number 1 and is equivalent to another
path $C^{(0)} (t_0)$ along the real-time axis plus the loop $C^{(1)} (0)$ of the winding number 1 encircling $z=0$ while the path $C^{(3)} (t_0)$ in Fig. 2 has
the winding number $3$. Thus, since the time integral along $C^{(0)} (t_0)$ in the real-time axis is unity,
the time integral along a curve $C(t_0)$ consisting of $C^{(0)} (t_0)$ and $C^{(n)} (0)$ of winding number $n$ around $z= 0$ becomes
\begin{eqnarray}
{\rm T}  \exp \bigl[-i \oint_{C (t_0)} \tilde{H} (z) dz \bigr] = {\rm T}  \exp \bigl[-i \oint_{C^{(n)} (0)} \tilde{H} (z) dz \bigr]
\end{eqnarray}

\begin{figure}[t]
{\includegraphics[width=0.475\linewidth]{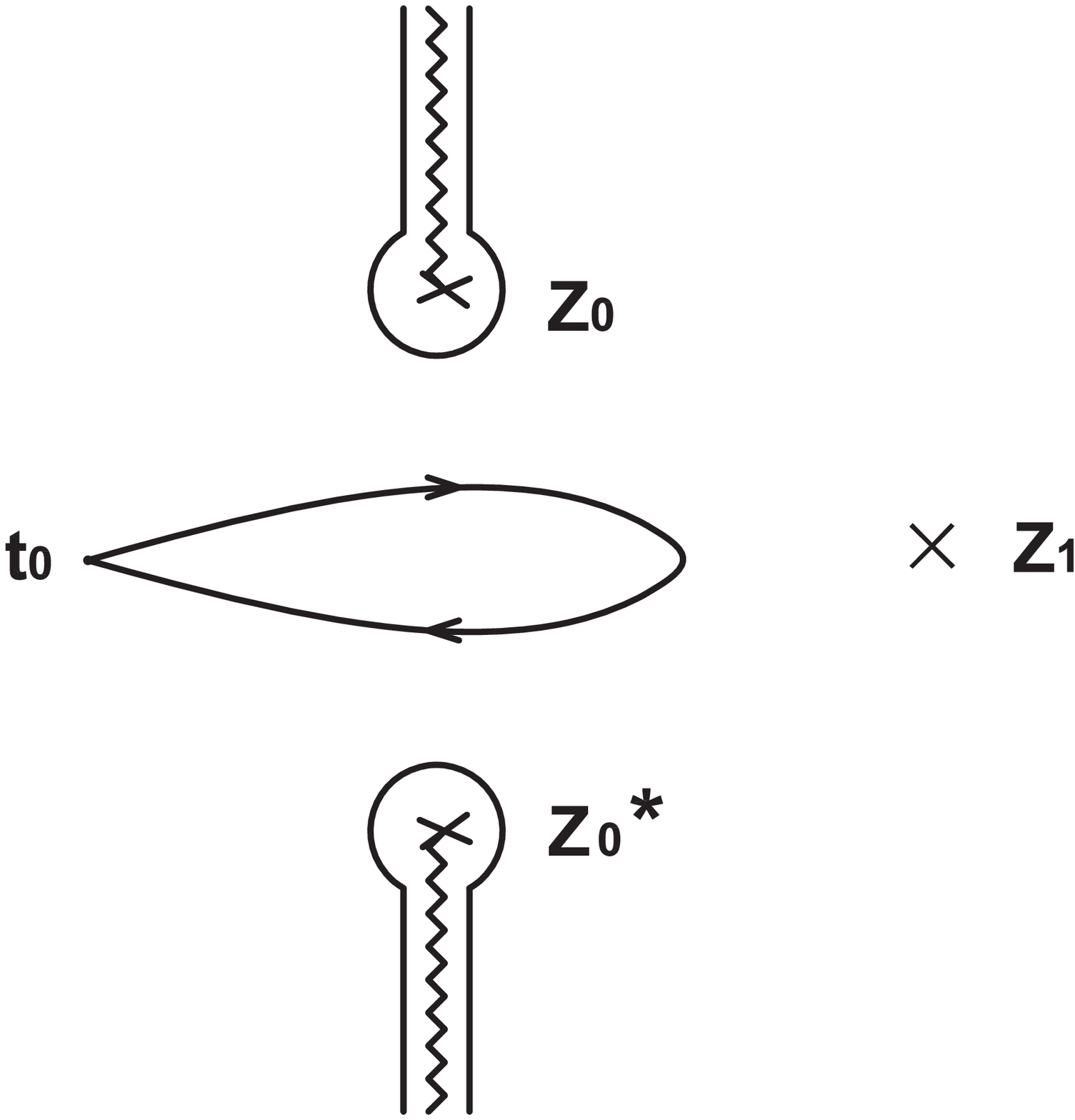}}\hfill
{\includegraphics[width=0.475\linewidth]{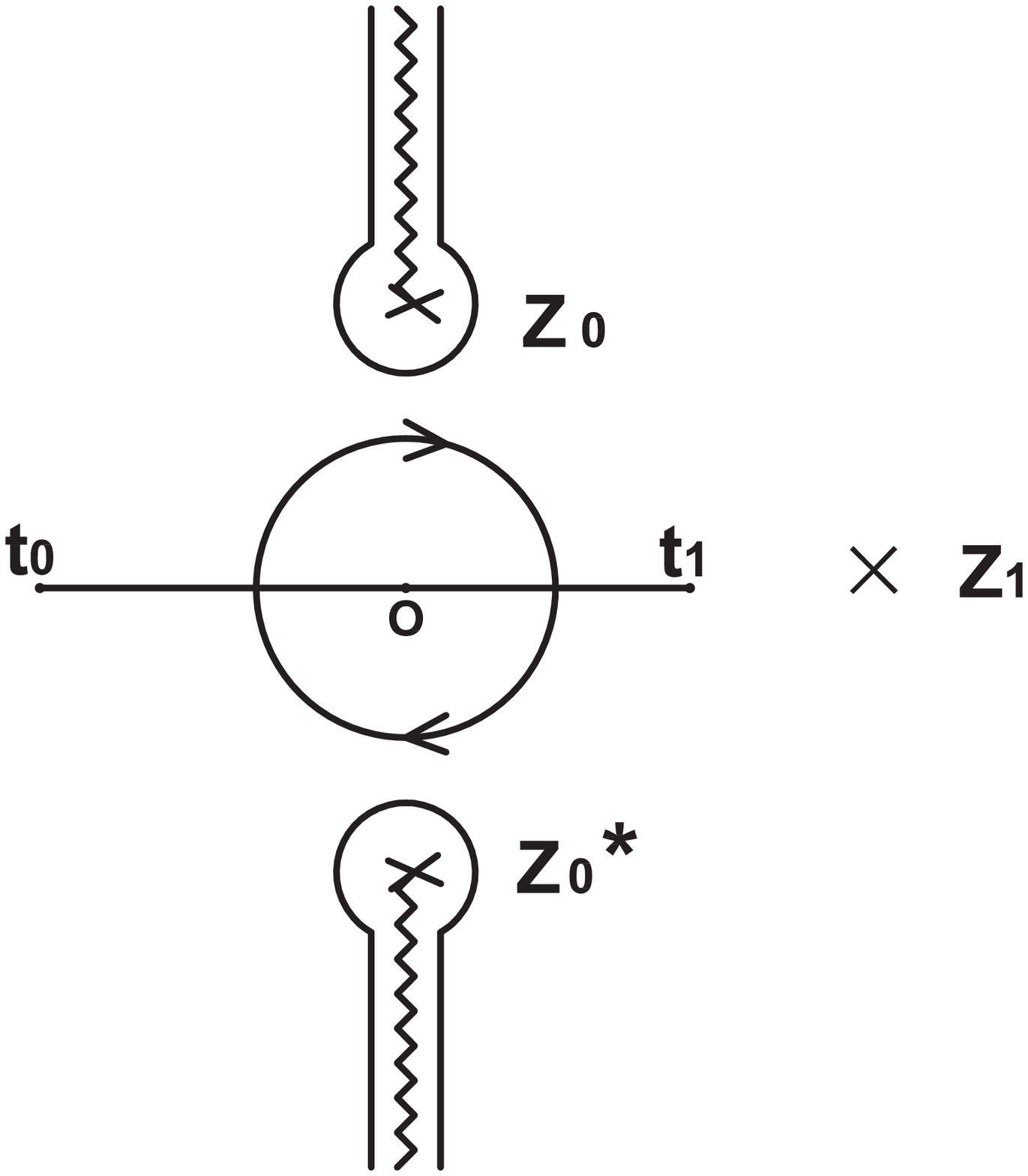}}
\caption{The frequency $\omega (z)$ has two branch points at $Z_0$ and $Z_0^*$, which are isolated by two branch cuts, and
has the simple pole located at $Z_1 = \infty$. The path $C^{(1)} (t_0)$ starts from the base point $t_0$, follows a loop clockwise and returns to $t_0$, excluding the simple pole
at the infinity [left panel]. The equivalent path consists of a real-line segment $C^{(0)} (t_0)$ from $t_0$ to $t_0$ and a loop $C^{(1)} (0)$ encircling $z =0$  [right panel].} \label{contour-1}
\end{figure}
\begin{figure}[t]
{\includegraphics[width=0.475\linewidth]{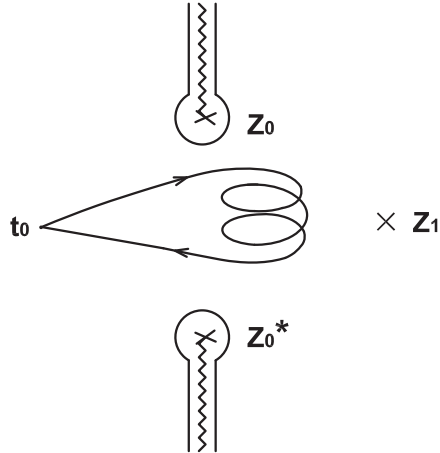}}
\caption{The path $C^{(3)} (t_0)$ starts from the base point $t_0$, follows clockwise a loop of winding number $3$ and returns to $t_0$. } \label{contour-2}
\end{figure}

However, in the lowest order of the Magnus expansion \cite{Magnus}, the residue theorem holds for the matrix-valued $\tilde{H}(z)$ \cite{Dollard-Friedman}
\begin{eqnarray}
{\rm T}  \exp \bigl[-i \oint_{C^{(n)} (0)} \tilde{H} (z) dz \bigr] = \exp \bigl[- 2 \pi n {\rm Res}[\tilde{H} (z_1)] \bigr], \label{res geom}
\end{eqnarray}
where ${\rm Res}[\tilde{H} (z_1)]$ is the residue of the simple pole $z_1$ at infinity \cite{Markushevich}. In fact, the frequencies (\ref{qed freq}) and (\ref{ds freq}) do have the simple pole at infinity, so the scattering amplitude between the in-vacuum and the transported in-vacuum is approximately given by
\begin{eqnarray}
\langle 0,  C^{(n)} (t_0) \vert 0, t_0 \rangle = e^{- \pi n {\rm Res}[\omega(z = \infty)]},
\end{eqnarray}
where the factor of $1/2$ for the vacuum state (\ref{h mat}) is taken into account and $n$ is the winding number.
The dynamical phase does not contribute to the scattering amplitude since it returns to $t_0$ and the frequency does not have any finite simple poles. The exponentially decaying scattering amplitude implies transitions to excited states, that is, particle production.
The residue at infinity is found by the large
$z$-expansion of the complex frequency (\ref{com freq})
\begin{eqnarray}
\omega (z) = f(z) \Bigl[ z - \frac{z_0 + z_0^*}{2} - \frac{(z_0- z_0^*)^2}{8 z} + \cdots \Bigr]. \label{residue}
\end{eqnarray}
In the first case of charged scalars in the constant field, the proper Riemann sheet is the entire complex plane with branch cuts as shown in Fig. 1.
Then the magnitude square of the scattering amplitude
\begin{eqnarray}
\vert \langle 0,  C^{(n)} (t_0) \vert 0, t_0 \rangle^2 \vert = e^{- n \pi \frac{m^2 + {\bf k}_{\perp}^2}{qE}}, \label{schwinger formula}
\end{eqnarray}
is the Schwinger pair-production rate for $n$-pairs of charged particles and antiparticles. It is analogous to the multi-instanton actions for pair production in the Coulomg gauge for static electric fields \cite{Kim-Page02}.
In the second case of real scalars in the dS space, the proper Riemann sheet $- \pi/H_{HC} < {\rm Im}~t \leq \pi/H_{HC}$ and a conformal mapping $e^{H_{HC}t} = z$ may be chosen, in which the scattering amplitude square
\begin{eqnarray}
\vert \langle 0, C^{(n)} (t_0) \vert 0, t_0 \rangle^2 \vert = e^{- 2 n \pi \frac{m}{H_{HC}}}, \label{ds formula}
\end{eqnarray}
is the Boltzmann factor for Gibbons-Hawking radiation in the dS space.

In summary, we showed that the geometric transition from a simple pole at infinity in the complex-time plane could explain particle production in a constant electric field and in a dS space. The Fourier-decomposed Hamiltonian for charged scalars in the constant electric field and for real scalars in dS space is infinite number of oscillators with time-dependent frequencies and/or mass. In the real-time evolution, any state prepared at an initial time that evolves into a future time and returns to the initial time remains in the same state with a trivial phase factor since there is no level-crossing, so the scattering amplitude between the in-vacuum and the transported in-vacuum is unity. However, we argued that the evolution along a closed path in the complex-time plane obtains a new geometric transition coming from the residue of the simple pole at the infinity, which differs from the geometric transition coming from level-crossings. Further, the scattering amplitude between the in-vacuum and the in-vacuum transported along a path of winding number $n$ leads to production of $n$-pairs.

Finally, a few comments are in order. First, it is worth to note that the geometric transition for the in-vacuum, though a consequence of nonstationarity of the Hamiltonian, resolves the factor of two puzzle for tunneling interpretation of Hawking radiation. The factor of two puzzle was explained in different ways \cite{Chowdhury,AAS06,AAS07,Nakamura} and by including the temporal contribution under the coordinate transformation from the embedding geometry \cite{APS,APGS08,APGS09}. The scattering amplitude between the in-vacuum and the transported in-vacuum takes the vacuum energy into account, which counts only a half of the energy quanta. Second, the massless limit of eq. (\ref{ds formula}) gives a unity scattering amplitude square between the in-vacuum and the transported in-vacuum in the complex plane as in eq. (\ref{real-time scat}). In fact, under the conformal mapping
$z = e^{H_{HC}t}$ the infinity has a double pole and thus does not contribute to the residue, which implies that the probability for the transported in-vacuum to remain in the in-vacuum is unity. The result is consistent with no production of massless particles in dS spaces in the in-out formalism, but there are subtle issues in the massless limit \cite{Akhmdeov-Buividovich,ABS}. Third, the geometric transition can be generalized to a frequency that has many level crossings and finite simple poles, which is the case of generic time-dependent vector potentials and global coordinates of dS spaces
and the Friedmann-Robertson-Walker spacetime \cite{ZRS}. Then the homotopy classes of paths are classified by finite simple poles inside the loops. The homotopy classes may have something to do with the Stokes phenomenon for particle production \cite{Dumlu-Dunne,Kim10}, which is beyond the scope of this paper. Another issue not pursued in this paper is the stimulated pair production from an initial particle state.

\acknowledgments
The author thanks Eunju Kang for drawing figures.
This paper was initiated during the Asia Pacific School on Gravitation and Cosmology (APCTP-NCTS-YITP Joint Program) in Jeju, Korea in 2013, benefited from helpful discussions at the Workshop on Gravitation and Numerical Relativity (APCTP Topical Program) and completed at Yukawa Institute for Theoretical Physics, Kyoto University. This work was supported by Basic Science Research Program through
the National Research Foundation of Korea (NRF) funded by the Ministry of Education (NRF-2012R1A1B3002852).

\end{document}